# Time-Domain Large Signal Investigation on Dynamic Responses of the GDCC Quarterly Wavelength Shifted Distributed Feedback Semiconductor Laser

Abdelkarim Moumen, Abdelkarim Zatni, Abdenabi Elyamani, Hamza Bousseta

M.S.I.T Laboratory, Department of Computer Engineering high school of technology, Ibnou Zohr University.

Abdelhamid Elkaaouachi

Department of Physics, Faculty of Sciences, Ibnou Zohr University.

*Abstract*—**A numerical investigation on the dynamic large-signal analysis using a time-domain traveling wave model of quarter wave-shifted distributed feedback semiconductor lasers diode with a Gaussian distribution of the coupling coefficient (GDCC) is presented. It is found that the single-mode behavior and the more hole-burning effect corrections of quarter wave-shifted distributed feedback laser with large coupling coefficient can be improved significantly by this new proposed light source.**

## I. INTRODUCTION

Long-haul modern Fiber-Optic Telecommunication Systems need optical source with high quality: high output optical power, low threshold current and reduced spatial hole burning effects, the longitudinal side mode are undesirable due to the presence of fiber dispersion [1][2][3][4]. The distributed feedback semiconductor lasers diode (DFB) have attracted great attention as the most favorable candidate. But the main disadvantage of this laser was the mode degeneracy and high threshold [1][6]. A phase shift along laser cavity can be introduced to remove the mode degeneracy [2]. Experimental results and numerical simulations have shown that the quarterly wavelength shifted distributed feedback laser (the phase shift is located at the center of the cavity and its value is fixed at $90^0$) oscillates at the Bragg wavelength, presenting the smallest threshold current and the high gain selectivity when compared to other $\lambda$ Phase-Shifted DFB laser diodes [1][2]. However, presences of the phase shift in the grating of DFB laser generally causes spatial no-uniformity and more interaction between the photon and carrier densities, especially for at high injection currents, this phenomenon, called spatial hole burning effect, reduce the performances of the $\lambda$ Phase-Shifted DFB lasers diodes [2][4]. Recently, $\lambda/4$ Phase-Shifted DFB with Gaussian distribution of the coupling coefficient (GDCC QWS) is proposed [1] to overcome the influence of spatial hole burning effect by maintaining uniform internal field along the laser cavity and reduce the threshold current, extensive studies have verified that stable single-mode and high power operation can be achieved in GDCC lasers with large coupling coefficient

$\kappa L = 2.5$, the studies is conducted by the Transfer Matrix Model (TMM) [1]. However, the relative important characteristics of this structure as the dynamic response have not been considered in their investigation.

In this letter the study consists in comparing the performance of the new proposed light source (GDCC QWS DFB) and conventional lasers having same total coupling coefficient in order to show the superiority of the GDCC configurations. The transient responses of the devices under analysis will be analyzed by using the time-domain multimode algorithm that is capable of including the longitudinal variation of the optical-mode and photon density profiles, the parabolic model of material gain is assumed [1]. In addition the spontaneous emission noise, the no uniform carrier density resulting from the hole burning effects as well as that the refractive index distribution are also taken into account, As a result this model may be applied to multi-sections lasers, such as phase-tunable lasers, tunable lasers and lasers designed to compensate for spatial hole burning (Subject of this paper). The model may also be applicable to tunable DFB laser amplifiers, the noise properties of DFB laser amplifiers and to bistable DFB switches.

The paper is organized as follows: the time-domain model is briefly described in section II. Simulation resultants of structures under analysis are presented and discussed in section III. Finally, a brief conclusion is drawn.

## II. TIME-DOMAIN MODEL (TDM)

For the phase-shifted distributed feedback semiconductor laser diode, the electric field in the laser cavity is given by [8]:

$$\psi(z,t) = [R(z,t)e^{-i\beta_0 z} + S(z,t)e^{+i\beta_0 z}]e^{-i\omega_0 t} \qquad (1)$$

Where $\beta_0$ is the propagation constant at Bragg frequency and $\omega_0$ is the reference frequency. $R$ and $S$ are the slowly varying complexes fields components include the amplitude and phase information of the forward and reverse wave in the waveguide respectively.

The fields $R$ and $S$ can be derived from the Maxwell's equations using the slowly varying amplitude approximation. Time-dependent coupled equation can be written as [8][11]:

$$\frac{1}{c_g}\frac{\partial R}{\partial t} + \frac{\partial R}{\partial z} =$$
$$\left[\left(\frac{1}{2}\left(\frac{\Gamma g(z,t)}{1+\varepsilon P} - \alpha_l\right) - i\delta\right)R + i\kappa(z)S\right]e^{-i\varphi(z)} + \xi(z,t) \quad (2)$$

$$\frac{1}{c_g}\frac{\partial S}{\partial t} - \frac{\partial S}{\partial z} =$$
$$\left[\left(\frac{1}{2}\left(\frac{\Gamma g(z,t)}{1+\varepsilon P} - \alpha_l\right) - i\delta\right)S + i\kappa(z)R\right]e^{+i\varphi(z)} + \xi(z,t) \quad (3)$$

Where $c_g$ is the group velocity, $\Gamma$ is the optical confinement factor, $\kappa(z)$ is the coupling coefficient between the forward and backward propagation waves, $\alpha_l$ is the waveguide loss (includes the absorption in both the active and cladding layer as well as any scattering), $\varphi(z)$ is the phase shift at $z$ position, $\varepsilon$ is the gain compression coefficient (non-linear coefficient to take into account saturation effects) and $\xi$ is the spontaneous emission term contributed to the forward and backward propagation components, the stochastic property of the noise term ($\xi$) is described by a random process with zero mean value and correlation function as described in [8][9][10] satisfying the correlation:

$$\begin{cases} \langle\xi(z,t),\xi^*(z',t')\rangle = \frac{\beta K B N^2}{c_g L}\delta(t-t')\delta(z-z') \\ \langle\xi(z,t),\xi(z,t)\rangle = 0 \end{cases} \quad (4)$$

Where $\beta$ is the spontaneous coupling factor, $K$ is the Peterman Coefficient and $\frac{BN^2}{c_g L}$ is the bimolecular recombination per unit length contributed to spontaneous emission.

$g(z,t)$ is the material gain, given by the parabolic formula:

$$g(z,t) = A_0\Delta N(z,t) - A_1[\Delta\lambda + A_2\Delta N(z,t)]^2 \quad (5)$$

In the above equation, $A_0$ is the differential gain, $A_1$ and $A_2$ are parameters used in the parabolic model assumed for the material gain, $\Delta N$ and $\Delta\lambda$ are the change of the carrier density and lasing wavelength defined as:

$$\begin{cases} \Delta N(z,t) = N(z,t) - N^0 \\ \Delta\lambda = \lambda - \lambda_0 \end{cases} \quad (6)$$

$N$ is the carrier density, $N^0$ is the carrier concentration at transparency ($g = 0$), $\lambda$ the oscillating wavelength and $\lambda_0$ is the peak wavelength at transparency. Using the first-order approximation for the refractive index, one obtains:

$$n(z,t) = n_0 + \Gamma\frac{dn}{dN}N(z,t) \quad (7)$$

Where $n_0$ is the refractive index at zeros carrier injection and $\frac{dn}{dN}$ is the differential index.

The $\delta$ in equations (2) and (3) represent the mode detuning (derivation from Bragg condition) defined as:

$$\delta(z,t) = \frac{2\pi}{\lambda}n(z,t) - \frac{2\pi n_g}{\lambda\lambda_B}(\lambda - \lambda_B) - \frac{\pi}{\Lambda(z)} \quad (8)$$

Where $\lambda_B$ is the Bragg wavelength, $\lambda$ is the lasing-mode wavelength, $n_g$ is the group refractive index and $\Lambda$ is the pitch (period) of the grating.

The carrier concentration $N(z,t)$ and the stimulated photon density are coupled together through the time-dependent carrier rate equation in the active layer which is shown here as [14]:

$$\frac{dN}{dt} = \frac{I}{qV} - \frac{N}{\tau} - BN^2 - CN^3 - \Gamma c_g g(z,t)\frac{P}{1+\varepsilon P} \quad (9)$$

Where $I$ is the injection current, $q$ is the modulus of the electron charge, $V$ is the volume of the active layer, $\tau$ is the carrier life time, $B$ is the radiative spontaneous emission coefficient and $C$ is the Auger recombination coefficient and $P$ is the photon density, which is related to the magnitude of travelling wave amplitudes as:

$$P(z,t) = |R(z,t)|^2 + |S(z,t)|^2 \quad (10)$$

In the TDM simulation the Large-signal spatiotemporal response of the laser is obtained by solving directly in the time domain the coupled wave equation (2)-(3) and the carrier rate equation (9) with axially-varying parameters. A finite-difference time-domain algorithm is applied to these equations with uniform intervals of time and space, to take the spatial hole burning and the carrier induced refractive index fluctuation into consideration, the laser cavity is divided into a large number of Subsections ($M = 5000$) with length $\Delta z = \frac{L}{M} = c_g\Delta t$, $L$ is the length of the cavity. In each section the material and structure parameters are kept constant, also the reflectivity at the end facet supposed to be zero. The numerical method followed here is similar to the one developed in [14].

The time-domain model is applicable to various types of semiconductor laser diodes. In this letter we apply the numerical model to compare to performance of the conventional quarterly wavelength shifted distributed feedback laser and GDCC QWS DFB laser having the same total coupling coefficient. In the proposed quarterly wavelength shifted distributed feedback laser the $\lambda$ phase-shifted is located at the centre of the cavity and the coupling coefficient $\kappa$ is a function of the longitudinal coordinate $z$, $\kappa$ change continuously along the laser cavity as follows:

$$\kappa(z) = \kappa_0 e^{-G\left(\left(z - \frac{L}{2}\right)/L\right)^2} \quad (11)$$

Where $\kappa_0$ the average value of the coupling coefficient, this parameter is introduced in order to allow a straightforward comparison between the characteristics of the GDCC QWS DFB and the conventional QWS DFB. The parameters definitions of these structures are summarized in table I, their distribution of the coupling coefficient are presented in the figure 1.

TABLE I.    SUMMARY PARAMETRS DEFINITIONS OF STRUCTURES

| Acronym | $G$ | $\kappa_0 L$ |
|---|---|---|
| Conventional QWS DFB | 0 | 2,50 |
| GDCC QWS DFB | 1 | 2,7098 |

TABLE II.    SUMMARY MATERIAL AND STRUCTURAL PARAMETRS

| SYMBOL | PARAMETRS | VALUE |
|---|---|---|
| $\tau$ | Carrier lifetime | $4.10^{-9} s$ |
| $B$ | Bimolecular recombination | $10^{-16} m^3 s^{-1}$ |
| $C$ | Auger recombination | $3.10^{-41} m^6 s^{-1}$ |
| $N^0$ | Transparency carrier density | $1,5.10^{24} m^{-3}$ |
| $\varepsilon$ | Non-linear gain coefficient | $1,5.10^{-23} m^3$ |
| $A_0$ | Differential gain | $2,7.10^{-20} m^6$ |
| $A_1$ | Gain curvature | $1,5.10^{19} m^{-3}$ |
| $A_2$ | Differential peak wavelength | $2,7.10^{-32} m^4$ |
| $\alpha_l$ | Internal absorption | $4.10^3 m^{-1}$ |
| $n_g$ | Group index | 3,7 |
| $c_g$ | Group velocity | $2,7.10^{-32} m^4$ |
| $L$ | Cavity length | $500\ \mu m$ |
| $D$ | Active layer thickness | $0,12\ \mu m$ |
| $w$ | Active layer width | $1,5\ \mu m$ |
| $V$ | Volume for active region | $90\ \mu m^3$ |
| $\Lambda$ | Grating period | $227,039\ \mu m$ |
| $\lambda_B$ | Bragg wavelength | $1550\ nm$ |
| $\lambda_0$ | Peak wavelength at transparency | $1565\ nm$ [1] |
| $\Gamma$ | Optical confinement factor | 0,35 |
| $\varphi$ | Phase shift | $90^0$ |
| $\Omega$ | Residue corrugation phase at left facet | $0^0$ |

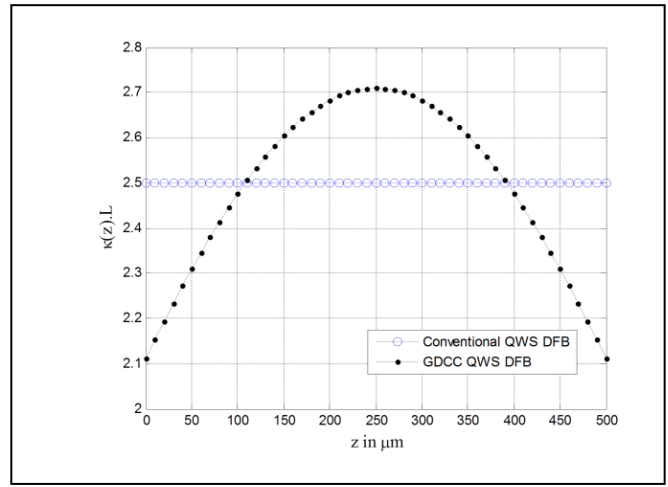

Figure 1.    Normalized coupling coefficient configurations used for the numerical simulations.

## RESULTS AND DISCUSSION

*Modest injection levels* $(I = 20mA)$

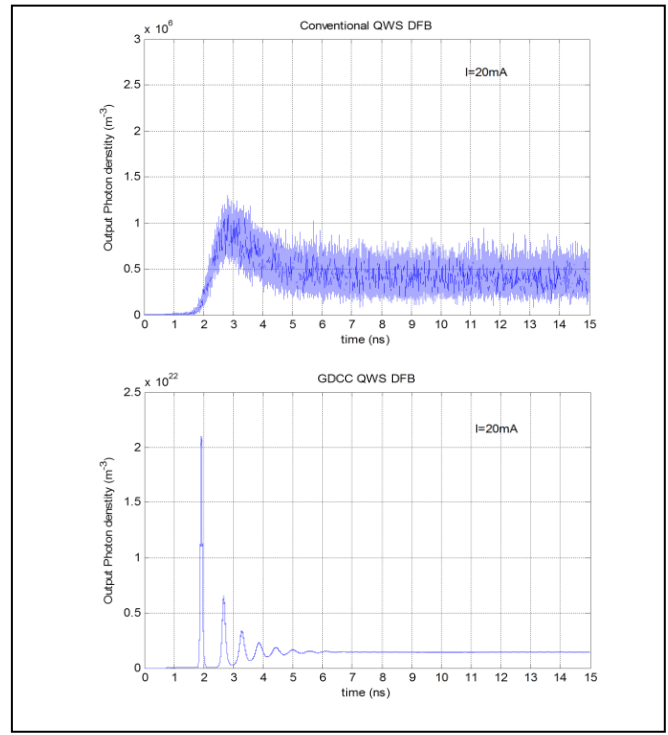

Figure 2.    Transient response(*current from* $0mA$ *to* $20mA$) of output photon density for the Conventional QWS DFB and the GDCC-QWS DFB.

When the biasing currents trends toward the threshold (Figure 2) the cavity is the seat of a spontaneous emission noise in the case of conventional QWS. Therefore this biasing current is inadequate to initiate the laser effects; the threshold current of the conventional structure is more than $20mA$, while the optical source has a low threshold current compared to the conventional QWS DFB, this first main advantage can be verified by evaluation of the optical power versus the current injection.

---

[1] According to the results obtained by the static study published in the [1]

From the emitting photon density at the facet, the output optical power can be evaluated. Figure 3 summaries results obtained for the conventional QWS DFB and GDCC QWS DFB LDs with the biasing current as parameter. Compared with the standard QWS DFB, it seems that the use of a smaller coupling coefficient near the facet has increased the overall cavity loss (case of GDCC QWS DFB Laser structure). The figure also shows that the GDCC QWS DFB laser structure has à relatively smaller value of threshold current $I_{th} = 19,75mA$ and a relatively larger output power under the same biasing current.

*High injection current ($I = 100mA$)*

In the figure 4, the damping of transient in GDCC QWS DFB is better than for the conventional device. After some relaxation oscillations, other differences occur between the conventional and GDCC QWS; the output photon density starts to oscillate in strong amplitude as the consequence of the beating between two modes in the case of conventional QWS DFB. This is confirmed by taking a sample of the emission spectrum in two different moments:

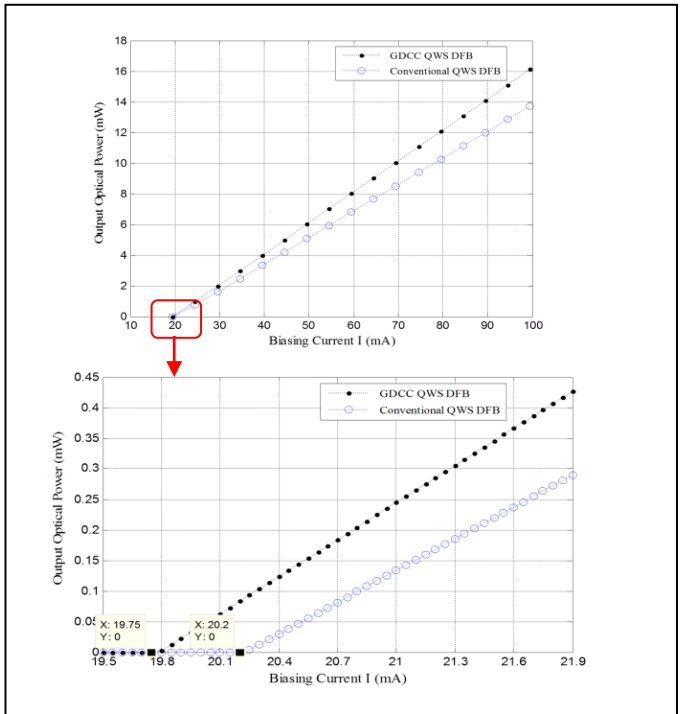

Figure 3. Emitted optical power versus current injection for the Conventional QWS DFB and the GDCC QWS DFB laser.

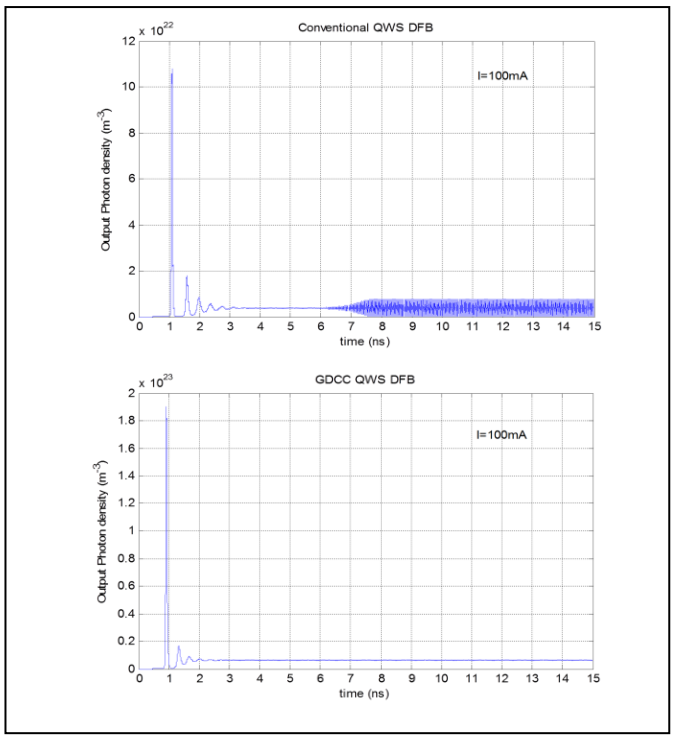

Figure 4. Transient response (*current from 0mA to 100mA*) of output photon density for the Conventional QWS DFB and the GDCC-QWS DFB.

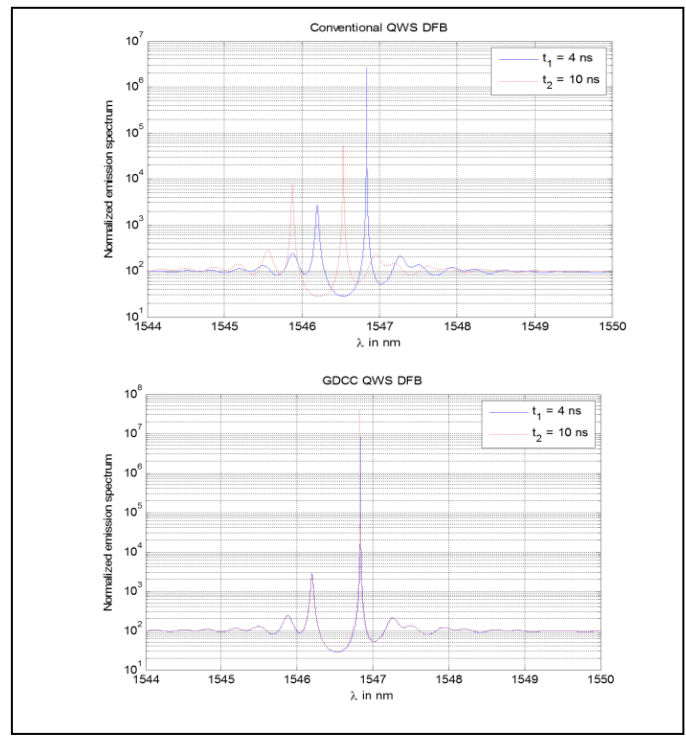

Figure 5. The Normalized emission spectrum in two different times, for the Conventional QWS DFB and the GDCC-QWS DFB.

turn-on-transient after the laser has switched and a typical oscillations are obtained in case of GDCC QWS DFB, this The spectral characteristics of the GDCC QWS DFB laser structures with the time changes are shown in the figure 5, distinct peaks which correspond to different oscillating modes are observed along the spectrum; the spectral amplitude of the dominant lasing mode found near $1546,90\ nm$ shows no sign of reduction and remains at a high value near $10^6$. Compared with the standard QWS DFB structure, the GDCC QWS DFB laser structure shows no server mode competition and an SMSR at least $25\ dB$ is maintained throughout of the time range.

In the case of the conventional QWS DFB, it can be seen that all peak wavelengths shift towards the shorter wavelength, and reduction of the spectral amplitude difference between the lasing mode and the side mode which is located at shorter wavelength side. At time $t_2 = 10\ n$, the side mode suppression ratio (SMSR) is reduced to less than $25\ dB$.

The variation of the longitudinal profiles of carrier density and refractive index can also indicate the occurrence of a multimode operation in DFB structures. As an illustration, we have plotted in the figure 5 the longitudinal profiles of refractive index in two distinct instants $(t_1, t_2)$ and the statistic longitudinal standard deviation of carrier density in the period $[t_1, t_2]$ given by:

$$\sigma(N_t) = \sqrt{E(N_t^2) - (E(N_t))^2} \qquad (12)$$

The beating between two modes observed in the case of conventional QWS DFB (Figure 5) is caused by the longitudinal hole burning effects. This phenomenon alters the lasing characteristics of the QWS DFB LD by changing the refractive index along the cavity (Figure 6 especially in the case of conventional structure). Under a uniform current injection, the light intensity inside the laser structure increases with biasing current. For strongly coupled laser devices, most light concentrate at the centre of the cavity. The carrier density at the centre is reduced remarkably as a result of stimulated recombination. Such a depleted carrier concentration induces an escalation of nearby injected carriers and consequently a spatially varying refractive index results Figure 6. This figure also shows the temporal instability of the carrier density especially near the facets of the cavity Conventional, which explains the strong amplitude oscillations observed for output photon density in the figure 4.

### III. Conclusion

With the help of a traveling wave model of semiconductor laser diodes, the dynamic analysis of Quarterly Wavelength Shifted Distributed Feedback Semiconductor Lasers with the Gaussian distribution of the coupling coefficient (GDCC) has been investigated and compared to conventional structures, to conduct this study we have developed a simple algorithm to calculate the large-signal dynamic response of DFB lasers by solving the time-dependent coupled wave equations and the carrier rate equation in the time domain. The spontaneous emission noise, longitudinal variations of carrier (hole burning) and photon densities as well as that of the refractive index are taken into consideration. The TDM was applied to GDCC

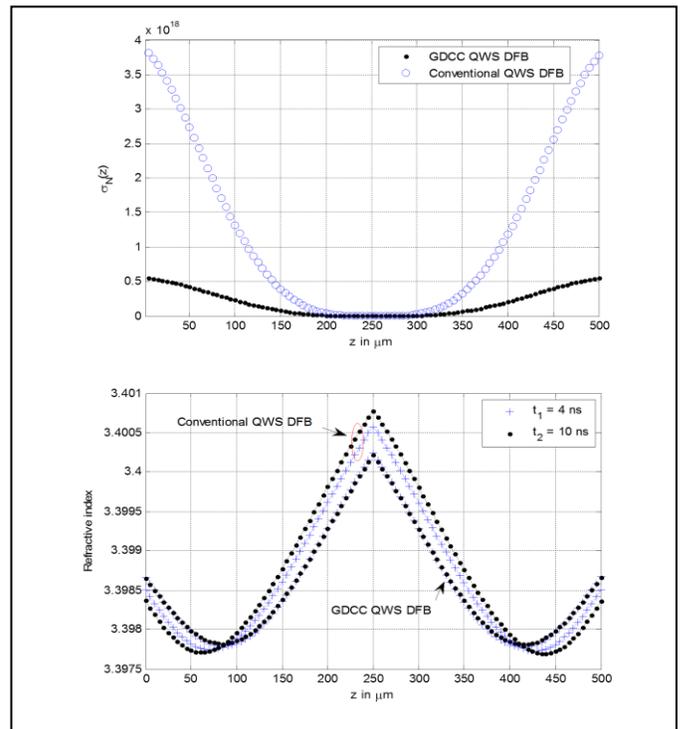

Figure 6. Statistic longitudinal standard deviation of carrier density in the period $[t_1, t_2]$ and the longitudinal distribution of refractive index in $t_1$ and $t_2$

QWS DFB and to Conventional QWS DFB, which is characterized by its uniform coupling coefficient, was shown to have a largest threshold current has the smallest output optical power. At high injection current, the conventional QWS structure is subject to mode beating and its output photon density starts to oscillate in strong amplitude as the result of the interference between the involved modes caused by the LSHB. Although the GDCC QWS DFB laser maintains a steady-state regime in which the output power becomes stabilized (no mode beating), no remarkable change in the spectral output in time, the damping of transient is better than for the conventional device. We may conclude that this new proposed light source can be used to extend the transmission distance in optical communication systems.

### Acknowledgment

This work was supported by the CNRST.

### References

[1]  A. Moumen, A. Zatni, A. Elkaaouachi, H. Bousseta, A. Elyamani, "A Novel Design of Quarter Wave-Shifted Distributed Feedback Semiconductor Laser for High-Power Single-Mode Operation," *Journal of Theoretical and Applied Information Technology* , vol. 38, No. 2, May 2012,

[2]  Ghafouri-shiraz, " Distributed feedback laser diodes and optical tunable filters," *Birmingham, UK: WILEY,* 2003

[3]  A. Zatni; J. Le Bihan, "Analysis of FM and AM responses of a tunable three-electrode DBR laser diode," *IEEE Journal of Quantum Electronics,* vol. 31, pp. 1009-1014, 1995.

[4] C. Ferreira Fernandes,"Hole-burning corrections in the stationary analusis of DFB laser diodes," *materials sciences & engineering B*, B74, pp. 75-79, 2000

[5] A.Zatni, "Study of the short pulse generation of the three quarter wave shift DFB laser (3QWS-DFB)," *Annals Of Telecommunications*, vol. 60, pp. 698-718, 2005

[6] Carlos. A. F. Fernandes; Jose B. M, Bovida "optimisation of an asymmetric three phase-shift distributed feedback semiconductor laser structure concerning the above-threshold stability, " *The European Physical Journal Applied Physics*, vol. 49, pp. 1-9, 2010

[7] T. Fessant, "Threshold and above-threshold analysis of corrugation-pitch-modulated DFB lasers with inhomogeneous coupling coefficient," *IEE Proc., Optoelectron*, vol. 144, pp. 365-376, 1997.

[8] A. Zatni, J. Le Bihan, A. Charaia and D. Khatib, "FM and AM responses of a three-electrode DBR laser diode, " *The 1st International Conference on Information & Communication Technologies: from Theory to Applications - ICTTA'04*, pp. 167-168, 2004

[9] Xin-Hong Jia, Dong-Zhong, Fei Wang, Hai-Tao Chen, "detailed modulation response analyses on enhanced single-mode QWS-DFB lasers with distributed coupling coefficient," *Optics communications* vol. 277, pp. 166-173, 2007

[10] L. M. Zhang, S. F. Yu, M. C. Nowell, D. D. Marcenac, J. E. Caroll, and R. G. S. Plumb, "Dynamic analysis of radiation and side-mode suppression in a second-order DFB Laser Using Time-domain large signal traveling wave model," *IEEE journal of quantum electronics*, vol. 30, No. 6, pp. 1389-1395, 1994.

[11] Jacques W. D. Chi, Lu Chao; M. K. Rao, "Time-Domain Large-Signal Investigation on Nonlinear interactions between An Optical Pulse and Semiconductor Waveguides," *IEEE Journal of Quantum Electronics*, vol. 37, No. 10, octobre 2001

[12] Thierry Fessant, "Enhanced Dynamics of QWS-DFB Lasers by Longitudinal Varaiation of their Coupling Coefficient," *IEEE photonics technology lettres*, vol. 9, No. 8, agust 1997

[13] Jing.-Yi. Wang and Michael Cada, "analysis and optimum design of distributed feedback lasers using coupled-power theory," *IEEE journal of quantum electronics*, vol. 36, pp. 52-58, 2000.

[14] A. Zatni, D. Khatib, M. Bour, J. Le Bihan "Analysis of the spectral stability of the three phase shift DFB laser (3PS-DFB), " *Annals of Telecommunications*, vol. 59, pp. 1031-1044, 2004

[15] F.Shahshahani,V.Ahmadim K. Mirabbaszadeh "concave tapered grating design of DFB laser at high power operation for reduced spatial hole-burning effect," *materials science and engineering* , vol. B96, pp. 1-7, 2002.

[16] Thierry Fessant, "Influence of a Nonuniform Coupling Coefficient on the Static and Large Signal Dynamic Behaviour of Bragg-Detnued DFB Lasers," *IEEE photonics technology lettres*, vol. 16, No. 3, March 1998

[17] G.-X. Chen, W. Li, C.-L.Xu, W.-P. Huang, S.-S. Jian, "Time and Spectral Domain Properties of Distributed Feedback-Type Gain Clamped Semiconductor Optical Amplifiers," *IEEE photonics technology lettres*, vol. 18, No. 8, April 2006

[18] M. G. Davis, R. F. O'Dowd, "A Transfer Matrix Method Based Large-Signal Dynamic Model For Multielectrode DFB Lasers," *IEEE Journal of Quantum Electronics*, vol. 30, No. 11, November 1994

AUTHORS PROFILE

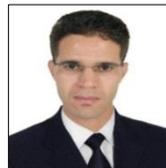

**Abdelkarim. MOUMEN** received the MSc degree in electrical and electronics system engineering from faculty of sciences University Ibnou Zohr in 2008; he is currently working the PhD at the centre of doctoral studies (Ibnou Zohr CED). His research interests include design, characterization, modelling and optimization of optoelectronic components and fibre optic communications systems.

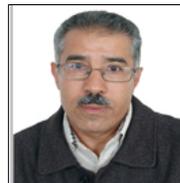

**Abdelkarim. ZATNI** was educated at the Telecom Bretagne University France; He obtained a PhD at the National School of Engineers of Brest France in 1994. He has been teaching experience for 20 years. He is currently a Professor and the Head of computer science department in Ibnou Zohr University at Higher School of technology Agadir, Morocco; He conducts his research and teaches in computer science and Telecommunications.